\documentclass[conference]{IEEEtran}
\IEEEoverridecommandlockouts

\usepackage{cite}
\usepackage{amsmath,amssymb,amsfonts}
\usepackage{algorithmic, booktabs}
\usepackage{graphicx}
\usepackage{textcomp}
\def\BibTeX{{\rm B\kern-.05em{\sc i\kern-.025em b}\kern-.08em
    T\kern-.1667em\lower.7ex\hbox{E}\kern-.125emX}}

\usepackage[table]{xcolor}
\usepackage{etoolbox}
\usepackage{pgf} 
\definecolor{high}{HTML}{76f013}  
\definecolor{low}{HTML}{ec462e}  
\newcommand*{\minval}{0.0}
\newcolumntype{C}[1]{@{}>{\columncolor{white}[0pt]\centering\arraybackslash}p{#1}@{}}

\usepackage{xcolor}
\newcommand{\gradientcell}[1]{
    \ifdimcomp{#1pt}{>}{100 pt}{#1}{
        \ifdimcomp{#1pt}{<}{0 pt}{#1}{
            \pgfmathparse{int(round(1*(#1/(1))-(\minval*(1/(1)))))}
            \xdef\tempa{\pgfmathresult}
            \cellcolor{green!\tempa!red!30} #1
    }}
}
    
\usepackage{acronym}
\usepackage{flushend}
\acrodef{PESQ}[PESQ]{Perceptual Evaluation of Speech Quality}
\acrodef{PEAQ}[PEAQ]{Perceptual Evaluation of Audio Quality}
\acrodef{POLQA}[POLQA]{Perceptual Objective Listening Quality Assessment}
\acrodef{DNN}[DNN]{Deep Neural Network}
\acrodef{MOS}[MOS]{Mean Opinion Score}
\acrodef{DNS}[DNS]{Deep Noise Suppression}
\acrodef{STOI}[STOI]{Short-Time Objective Intelligibility}
\acrodef{ESTOI}[ESTOI]{Extended STOI}
\acrodef{RVQ}[RVQ]{Residual Vector Quantization}
\acrodef{SQR}[SQR]{Signal-to-Quantization error power Ratio}
\acrodef{LQR}[LQR]{Latent-representation-to-Quantization error power Ratio}
\acrodef{ODAQ}[ODAQ]{Open Datasets of Audio Quality}
\acrodef{SNR}[SNR]{Signal-to-Noise power Ratio}
\acrodef{SIR}[SIR]{Signal-to-Interferer power Ratio}
\acrodef{SAR}[SAR]{Signal-to-Artifact power Ratio}
\acrodef{SDR}[SDR]{Signal-to-Distortion power Ratio}
\acrodef{NMR}[NMR]{Noise-to-Mask Ratio}

\begin{document}

\title{On the Relation Between Speech Quality and Quantized Latent Representations of Neural Codecs 
}

\author{Mhd Modar Halimeh, Matteo Torcoli, Philipp Grundhuber, and Emanu\"el A. P. Habets \\
\textit{Fraunhofer Institute for Integrated Circuits IIS, Erlangen, Germany}
}

\maketitle

\begin{abstract}
    Neural audio signal codecs have attracted significant attention in recent years. In essence, the impressive low bitrate achieved by such encoders is enabled by learning an abstract representation that captures the properties of encoded signals, e.g., speech. In this work, we investigate the relation between the latent representation of the input signal learned by a neural codec and the quality of speech signals. To do so, we introduce Latent-representation-to-Quantization error Ratio (LQR) measures, which quantify the distance from the idealized neural codec's speech signal model for a given speech signal. We compare the proposed metrics to intrusive measures as well as data-driven supervised methods using two subjective speech quality datasets. This analysis shows that the proposed LQR correlates strongly (up to 0.9 Pearson's correlation) with the subjective quality of speech. Despite being a non-intrusive metric, this yields a competitive performance with, or even better than, other pre-trained and intrusive measures. These results show that LQR is a promising basis for more sophisticated speech quality measures. 
\end{abstract}

\vspace*{-1mm}
\section{Introduction}
For decades, researchers have been investigating the perception of speech quality and the prediction thereof \cite{hu2007evaluation}. 
Speech quality perception spans several axes, such as the perception of distortions with or without comparison with a reference signal, as well as timbre, intelligibility, and listening effort.  

Numerous objective metrics have been proposed to estimate the perceived quality of audio signals \cite{hu2007evaluation, Torcoli2021}. 
These include intrusive measures, i.e., using a reference signal (e.g., \cite{pesq, stoi, ESTOI, twofmodel, brandenburg1987evaluation, PEAQ, bsseval}), and non-intrusive or reference-free metrics (e.g., \cite{wvmos, dnsmos}).


In parallel, significant progress has been achieved in neural audio compression, enabling high-fidelity audio signal reconstruction at impressively low bitrates. For instance, LPCNet \cite{LPCNet} combines hand-crafted features and uniform quantization to realize an efficient vocoder. Meanwhile, the authors in \cite{VQVAEWAVENET} proposed conditioning a WaveNet vocoder on a VQ-VAE discrete latent representation. More recently, many proposals following the approach proposed in \cite{soundstream} can be found in the literature \cite{audiodec, funcodec, encodec}. This approach can be summarized by two components: a \ac{DNN}-based autoencoder (e.g., fully convolutional as in \cite{soundstream}) and a quantization layer separating the encoder and decoder. Nonetheless, it is worth noting that in addition to their use in audio coding, discrete representations of audio signals have been widely used, e.g., speech \cite{directS2S} and music generation \cite{MUSICLM}. 

These advances in learning, potentially compressed, representations of audio signals motivated their use for quality estimation. Recently, several proposals utilize pre-learned latent space representations \cite{fu2024selfsupervised,Manocha:2020:ADP,ragano2024scoreq} to capture the quality of a given signal. For instance, \cite{fu2024selfsupervised} proposes to train a neural codec for clean speech signals only such that the quantization error can then be used directly to capture the quality of a speech signal. Meanwhile, the authors in \cite{ragano2024scoreq} utilize contrastive learning to, e.g, fine-tune a pre-trained wav2vec2 \cite{baevski2020wav2vec20frameworkselfsupervised} model.


In this paper, we hypothesize a relation between speech quality and trainable quantization used to obtain a discrete latent representation in neural codecs. To study this relation, we propose two \acp{LQR}, which are computed using pre-trained neural codecs without additional training conditions (e.g., as done in \cite{fu2024selfsupervised}) or mapping. We then propose two \ac{LQR}-based metrics and show that these strongly correlate with the perceived speech quality. Consequently, \ac{LQR} enables speech quality metrics that are non-intrusive and require no training using large datasets of subjective listening tests.
Furthermore, as shown by the analysis using subjective listening tests, the proposed \ac{LQR}-based metrics are competitive with, or even outperform, other pre-trained and intrusive measures. 


\vspace*{-1mm}
\section{Neural Audio Coding} \label{sec:proposed_method}
Most neural codecs consist of an encoder-decoder architecture, and a quantization layer at the bottleneck \cite{soundstream}. The quantization layer can be realized through residual vector quantization, where the codec network is then adapted end-to-end to minimize a reconstruction loss at the decoder output, which may be measured by, e.g., an adversarial network. Furthermore, to allow for a flexible and accurate representation of the audio signal, the codebooks of the quantization layer are often optimized to minimize the quantization error introduced to the encoder output. 

In particular, a signal $s$ is processed by the encoder as 
\begin{equation}\vspace*{-1mm}
    x = \textrm{Encoder}(s). \label{eq:encoder}
\end{equation}
The latent representation is then quantized, e.g., iteratively employing \ac{RVQ}, where for the $k$-th quantization step
\begin{equation}
   q_k = \textrm{VQ}\left(x - \sum_{i=0}^{k-1} q_i \right); \quad q_0=0, \label{eq:quantization} \vspace{-1mm}
\end{equation}
where $\textrm{VQ}$ denotes the vector quantization operator \cite{vq}, while $q_k$ denotes the output of the $k$-th vector quantizer. The quantized representation is used, e.g., by concatenating the $K$ quantized tensors, to reconstruct the input signal $s$ such that 
\begin{equation}
    \hat{s} = \textrm{Decoder}(q_1, \dots, q_K). 
\end{equation}


\section{\acf{LQR}}

In this paper, we focus on speech signals. Consequently, the encoding of the input signal in Eq.~\eqref{eq:encoder} is viewed as processing the input signal by a speech signal model learned by the encoder, producing a corresponding set of feathers that are subsequently discretized by a quantizer.  


In other words, limiting the scope of the discussion to \ac{RVQ}, we assume that the $k$-th step quantization error in Eq.~\eqref{eq:quantization}
\begin{equation}
    e_k = x - \sum_{i=1}^{k} q_i,
\end{equation} 
has a variance $\sigma(e_k)$ that is calculated per frame and averaged over the number of signal frames. Given a codec trained to minimize the quantization error,  $\sigma(e_k)$ is proportional to the frequency (in the probabilistic sense) $f(x)$ observed during training, i.e., {$\sigma(e_k) \propto \frac{1}{f(x)}$}. 
%
The frequency $f(\cdot)$ is mainly determined by the signal model learned by the encoder and by the dataset used to train the network. 

Driven by empirical observations where neural coders often perform best for clean speech signals, we further assume that clean speech signals strongly determine the signal model. Consequently, clean speech signals are expected to result in features with a higher frequency $f(x)$ (or equivalently lower $\sigma(e_k)$) than noisy or degraded signals. 

Inspired by the conventional \ac{SQR} metric, we investigate two metrics to quantify the quantization error of the latent space representation:
\begin{itemize}
    \item $\overline{\textrm{LQR}}$: describes the average \ac{LQR}. It considers quantization errors resulting from each quantization step and is calculated as 
    $$
        \overline{\textrm{LQR}} = \frac{1}{K} \sum_{k=1}^{K} \textrm{LQR}_k,
    $$
    where $\textrm{LQR}_k$ denotes the \ac{LQR} of the $k$-th quantization step and is obtained by 
    $$
        \textrm{LQR}_k = \frac{ \sigma(e_{k-1} ) }{ \sigma(e_k)  },
    $$
    where $e_0=x$. Consequently, $\overline{\textrm{LQR}}$ includes the quantization errors introduced by the different quantization steps. 
    \item Alternatively, $\textrm{LQR}_{0, K}$ describes the input-to-$K$-th-quantization error power ratio. It differs from $ \textrm{LQR}_k $ in that it does not consider individual quantization steps but measures \ac{LQR} as the ratio between its input and last, i.e., $K$-th block, quantization error signal, 
    $$
        \textrm{LQR}_{0, K} = \frac{ \sigma(x) }{ \sigma(e_K)  }.
    $$
    
\end{itemize}

Either variant renders a higher \ac{LQR} values for signals with higher quality and vice versa.  Moreover, other variations of \ac{LQR}-based metrics are possible, e.g., by considering the median or geometric mean of the intermediate \ac{LQR} values. However, we exclusively examine the two aforementioned measures in the remainder of this paper as they are found experimentally to perform best.

\vspace*{-1mm}\section{Experimental Setup}

\subsection{Neural Encoders}
The following encoders were considered as the basis to calculate the \ac{LQR}.

\subsubsection{Encodec}

Encodec \cite{encodec} uses a sampling rate of $48$~kHz and is realized as a causal model. Five pre-trained models are considered with bitrates \sloppy{$\{1.5, 3.0, 6.0, 12.0, 24.0 \}$~kbps}. These models are denoted in the following by Encodec$_{\textit{bitrate}}$ where $\textit{bitrate}$ denotes the bitrate. As described in \cite{encodec}, the models are trained using a diverse dataset, which includes clean speech, noisy speech, music, and general audio signals. 

\subsubsection{Audiodec}
Audiodec \cite{audiodec} utilizes a different two-stage training strategy than Encodec. We limit the scope of our experiments to four pre-trained Audiodec models. These models use symmetric autoencoder architectures and are trained on clean-speech-only datasets, namely, VCTK \cite{VCTK} or LibriTTS \cite{LibriTTS}. The models operate at a $48$~kHz sampling rate, and they are denoted by $\textrm{Audiodec}_{\textit{dataset}}$. 

\subsubsection{Funcodec}
Funcodec \cite{funcodec} uses a similar architecture to Encodec. However, unlike Audiodec and Encodec, Funcodec supports multiple bitrates using structured quantization dropout. We utilize two model types (operating at $24$~kHz sampling rate), one trained on a speech-only dataset (denoted by \textit{academic}) trained using the LibriTTS dataset \cite{LibriTTS}, and a general purpose model (denoted by \textit{general}) trained on a speech-only (English and Chinese) dataset.

For brevity, out of the several pre-trained models considered per codec, only the results of the best-performing model (per codec) are presented. 

\vspace*{-1mm}
\subsection{Subjective Quality Scores}
This section describes the subjective quality scores we consider to be the ground truth in our experiments. Two independent, openly available datasets are used.

\subsubsection{The ODAQ Dataset}
The \ac{ODAQ} dataset \cite{Torcoli2024ODAQ} is an openly available dataset of audio samples and corresponding quality scores. The dataset was designed to provide a common benchmark for research into objective quality metrics. For this reason, it spans an extensive range of quality levels and distortion types. The subjective scores are obtained through MUSHRA~\cite{MUSHRA} listening tests.
To assess the correlation between \ac{LQR} metrics and speech quality, only items containing speech were used: This part of the dataset is referred to as DE in \cite{Torcoli2024ODAQ}, and includes movie-like stereo signals where speech recorded in a studio is mixed with different types of music and effects. Five speech enhancement systems are evaluated in the listening test.

\subsubsection{The CHiME-7 UDASE Dataset}\
The evaluation of the UDASE task of the CHiME challenge \cite{LEGLAIVE2025} offers an openly available collection of speech signals at different quality levels and corresponding subjective ratings. In this case, the subjective ratings result from a P.835 listening test \cite{P835}.
This provides an interesting additional benchmark, as P.835 differs significantly from MUSHRA employed in ODAQ. In fact, P.835 uses a different rating scale, focuses on speech, and listeners are asked to evaluate not only overall quality (OVRL) but also speech distortion (SIG) and intrusiveness of background noise (BAK). Furthermore, while MUSHRA employs expert listeners, P.835 listening test should involve exclusively non-expert assessors. Finally, P.835 does not include a reference or anchor signals. The usage of the rating scale can be somehow anchored by the training session, which should precede the first session of the listening test, and it should contain the full range of OVRL, SIG, and BAK levels. In \cite{LEGLAIVE2025}, the training session included the conditions described in \cite{etsi103}. The design of these training conditions is intended to independently modulate SIG, BAK, and OVRL over the entire rating range. This provides subsets of signals for which the speech quality was distorted without introducing any background noise (SIG only), and signals for which background noise was introduced without distorting the signal quality (BAK only). 
In addition, we consider the in-domain unlabeled eval set of~\cite{LEGLAIVE2025}, which consists of recordings made during real dinner parties featuring speakers in noisy reverberant rooms. Four single-channel speech enhancement systems are evaluated. 

\vspace*{-1mm}
\section{Experimental Results}
In this section, the \ac{LQR} metrics are investigated, and their performance as speech quality predictors is analyzed in terms of Pearson's linear correlation coefficient $\rho$ and Spearman's rank correlation coefficient $\tau$.
\ac{LQR} is compared to these widely-used intrusive and non-intrusive measures from the literature: \ac{PESQ}~\cite{pesq}, \ac{STOI} \cite{stoi},  \ac{ESTOI} \cite{ESTOI}, the \mbox{2f-model}~\cite{twofmodel}, \mbox{DNSMOS}~\cite{dnsmos}, \mbox{Wav2Vec2-MOS}~\cite{wvmos}, \ac{SIR}, \ac{SAR}, \ac{SDR} \cite{bsseval}, and 
\ac{NMR}~\cite{brandenburg1987evaluation} (as implemented in PEAQ \cite{PEAQ, Kabal}).


\begin{table}
    \caption{Correlation coefficients ($\rho$ and $\tau$ multiplied by 100) between objective metrics and average quality scores in \ac{ODAQ} (DE). The color map ranges from green (for large correlation coefficients) to red (for small correlation coefficients). }
    \centering
    \begin{tabular}{c c  c c c}
        \toprule
         Metric     & & Pearson's ($\rho$) &   & Spearman's ($\tau$)\\
         \midrule
         \ac{PESQ}              &           &   \gradientcell{80}    &    &  \gradientcell{69}    \\
         \ac{STOI}              &           &   \gradientcell{40}    &    &  \gradientcell{42}    \\ 
         \ac{ESTOI}             &           &   \gradientcell{51}    &    &  \gradientcell{49}    \\ 
         2f-model               &           &   \gradientcell{66}    &    &  \gradientcell{48}    \\ 
         NMR                    &           &   \gradientcell{89}    &    &  \gradientcell{87}    \\
         SIR                    &           &   \gradientcell{70}    &    &  \gradientcell{69}    \\ 
         SDR                    &           &   \gradientcell{2}    &    &  \gradientcell{23}    \\ 
         SAR                    &           &   \gradientcell{19}    &    &  \gradientcell{14}    \\ 
         Wav2Vec2-MOS            &           &   \gradientcell{49}    &    &  \gradientcell{45}    \\ 
         DNSMOS (BAK)          &           &   \gradientcell{43}    &    &  \gradientcell{38}     \\ 
         DNSMOS (SIG)          &           &   \gradientcell{20}    &    &  \gradientcell{17}     \\ 
         DNSMOS (OVRL)         &           &   \gradientcell{45}    &    &  \gradientcell{43}     \\ 
         \midrule
        \multicolumn{5}{l}{Audiodec$_\textrm{VCTK}$}                                         \\
        \midrule
$\overline{\textrm{LQR}}$       &           &   \gradientcell{56}    &    &  \gradientcell{55}   \\ 
$\textrm{LQR}_{0, K} $          &           &   \gradientcell{56}    &    &  \gradientcell{56} \\ 
\midrule
        \multicolumn{5}{l}{Audiodec$_\textrm{librTTS}$}                                         \\
\midrule
$\overline{\textrm{LQR}}$       &           &   \gradientcell{53}    &    &  \gradientcell{51}   \\ 
$\textrm{LQR}_{0, K} $          &           &   \gradientcell{53}    &    &  \gradientcell{52} \\ 
         \midrule
\multicolumn{5}{l}{Encodec$_{1.5}$}                                                  \\
        \midrule
$\overline{\textrm{LQR}}$       &          &   \gradientcell{69}    &      &  \gradientcell{71}  \\ 
$\textrm{LQR}_{0, K} $          &          &   \gradientcell{68}    &      &  \gradientcell{71}  \\ 
         \midrule
\multicolumn{5}{l}{Funcodec General}                                                   \\
        \midrule
$\overline{\textrm{LQR}}$       &         &    \gradientcell{11}   &       &  \gradientcell{13} \\ 
$\textrm{LQR}_{0, K} $          &         &    \gradientcell{23}   &       &  \gradientcell{21} \\ 
         \midrule
\multicolumn{5}{l}{Funcodec Academic}                                                   \\
        \midrule
$\overline{\textrm{LQR}}$       &         &    \gradientcell{17}   &       &  \gradientcell{20} \\ 
$\textrm{LQR}_{0, K} $          &         &    \gradientcell{20}   &       &  \gradientcell{24} \\ 
         \bottomrule
    \end{tabular}
    \label{tab:ODAQ_CC}
    \vspace{-4mm}
\end{table}

\vspace*{-1mm}
\subsection{Results on ODAQ}
As shown in Table~\ref{tab:ODAQ_CC}, the largest correlation coefficients are reported for the \ac{NMR} with $\rho=0.89$ and $\tau=0.87$.  \ac{PESQ} also performs competitively with $\rho=0.8$ and $\tau=0.69$, respectively. Meanwhile, objective signal measures such as \ac{SDR} perform poorly, as reflected by their small correlation coefficients, and also noted in previous works, e.g., \cite{Torcoli2021}. 

Meanwhile, the considered non-intrusive measures, i.e., the DNSMOS (BAK and OVRL) and Wav2Vec2-MOS, render moderate performance with the Wav2Vec2-MOS being the better metric with $\rho=0.49$ and $\tau=0.45$. 

As for the proposed $\overline{\textrm{LQR}}$ metric, we observe a strong dependency on the underlying audio encoding model. In particular, while Funcodec (in both variants) renders small correlation coefficients ($\leq0.24$), Encodec and Audiodec perform significantly better with correlation coefficients up to $0.71$. More specifically, Encodec$_{1.5}$ renders the best performance, with the correlation coefficients $\rho = 0.69$ and $\tau = 0.71$. This is notable, as Encodec is trained using a general audio dataset in contrast to, e.g., Audiodec, which is trained using speech signals only. This places the $\overline{\textrm{LQR}}$ as the best non-intrusive measure, even outperforming some intrusive measures, e.g., the 2f-model. In comparison, Audiodec$_{\textrm{VCTK}}$ performs worse than Encodec, however, achieving correlation coefficients, $\rho=0.53$ and $\tau=0.51$, that are larger than those of the considered non-intrusive DNSMOS and Wav2Vec2-MOS. 

Finally, comparing the two proposed metrics, i.e., the $\overline{\textrm{LQR}}$ and $\textrm{LQR}_{0, K}$, we can observe that the two metrics render almost identical performance, suggesting a significant dependency between speech signal quality and latent quantization errors regardless of the particularity of the \ac{LQR} measure.

{\setlength{\tabcolsep}{2pt}%
\begin{table*}[!t]
\caption{CHiME-7 UDASE Dataset. For improved compactness, $\rho$ and $\tau$ are reported multiplied by 100. Entries are left empty if it is not possible to compute intrusive measures for the lack of reference signals.}
    \centering
    \begin{tabular}{c l | c c   c c  c c | c c  c c  c c | c c  c c  c c }
    \toprule
         &   & \multicolumn{6}{c}{Training: SIG only} & \multicolumn{6}{c}{Training: BAK only} & \multicolumn{6}{c}{In-domain unlabeled eval} \\ 
    \midrule
    &  & \multicolumn{2}{c}{BAK} &  \multicolumn{2}{c}{SIG} &\multicolumn{2}{c}{OVRL} & \multicolumn{2}{c}{BAK} &   \multicolumn{2}{c}{SIG} &   \multicolumn{2}{c}{OVRL} & \multicolumn{2}{c}{BAK}    & \multicolumn{2}{c}{SIG} &   \multicolumn{2}{c}{OVRL} \\
    \midrule
    metric &   & $\rho$ & $\tau$    & $\rho$ & $\tau$    &  $\rho$ & $\tau$  %
                         & $\rho$   & $\tau$   & $\rho$ & $\tau$     & $\rho$ & $\tau$ %
                         & $\rho$   & $\tau$   & $\rho$ & $\tau$     & $\rho$ & $\tau$ \\
    \midrule
    PESQ &               & \gradientcell{37}  & \gradientcell{47}     & \gradientcell{91}  & \gradientcell{96}       & \gradientcell{94} & \gradientcell{97}  %
                         & \gradientcell{93}  & \gradientcell{96}     & \gradientcell{50}  & \gradientcell{86}       & \gradientcell{84} & \gradientcell{96}   %
                         & & & & & & \\   
    STOI &   & \gradientcell{40}   & \gradientcell{38}  &    \gradientcell{86} & \gradientcell{80}        & \gradientcell{89} & \gradientcell{82}  %
                         & \gradientcell{78}  & \gradientcell{96}    & \gradientcell{96} & \gradientcell{83}        & \gradientcell{89} & \gradientcell{95} %
                         & & & & & & \\
    ESTOI&   & \gradientcell{44} & \gradientcell{41}   & \gradientcell{87}   & \gradientcell{82}    &  \gradientcell{89} & \gradientcell{83}  %
                        & \gradientcell{85} & \gradientcell{95}   & \gradientcell{90}   & \gradientcell{80}    & \gradientcell{93}  & \gradientcell{94} %
                         & & & & & & \\
2f-model&   & \gradientcell{48} & \gradientcell{42}  & \gradientcell{88}   & \gradientcell{86}   &  \gradientcell{88} & \gradientcell{87}  %
                        & \gradientcell{73} & \gradientcell{87}  & \gradientcell{96}   & \gradientcell{83} & \gradientcell{86}  & \gradientcell{87} %
                         & & & & & & \\
    NMR&   & \gradientcell{28} & \gradientcell{46}   & \gradientcell{81}   & \gradientcell{94}   &  \gradientcell{85} & \gradientcell{94}  %
                        & \gradientcell{98} & \gradientcell{95}   & \gradientcell{63}   & \gradientcell{81}    & \gradientcell{92}  & \gradientcell{95} %
                         & & & & & & \\
\ac{SIR} &  & \gradientcell{54} & \gradientcell{46}   & \gradientcell{89} & \gradientcell{90}    &  \gradientcell{90} & \gradientcell{93}  %
                         & \gradientcell{99} & \gradientcell{97}  & \gradientcell{71} & \gradientcell{89}   & \gradientcell{96} & \gradientcell{97} %
                         & & & & & & \\
\ac{SDR}&   & \gradientcell{34} & \gradientcell{53}  & \gradientcell{86} & \gradientcell{95}   &  \gradientcell{90} & \gradientcell{97}  %
                         & \gradientcell{99}   & \gradientcell{99} & \gradientcell{76} & \gradientcell{88}   & \gradientcell{98} & \gradientcell{99} %
                         & & & & & & \\                   
\ac{SAR}&   & \gradientcell{36} & \gradientcell{54}   & \gradientcell{88} & \gradientcell{96}    &  \gradientcell{91}& \gradientcell{97}  %
                           & \gradientcell{95}   & \gradientcell{96}& \gradientcell{53} & \gradientcell{83}   & \gradientcell{87} & \gradientcell{96} %
                         & & & & & & \\
Wav2Vec2-MOS&   & \gradientcell{34} & \gradientcell{39}   & \gradientcell{80} & \gradientcell{71}    &  \gradientcell{83} & \gradientcell{71}%
                           & \gradientcell{92}   & \gradientcell{97}      & \gradientcell{90}  & \gradientcell{85}   & \gradientcell{97} & \gradientcell{96} %
                            & \gradientcell{65}   & \gradientcell{64}     & \gradientcell{30}  & \gradientcell{32}    & \gradientcell{53} & \gradientcell{55}  \\
DNSMOS (BAK)&              & \gradientcell{48}   & \gradientcell{43}     & \gradientcell{74}  & \gradientcell{71}   &  \gradientcell{75} & \gradientcell{73}  %
                            & \gradientcell{92}   & \gradientcell{95}     & \gradientcell{80}  & \gradientcell{84}   & \gradientcell{94} & \gradientcell{94} %
                            & \gradientcell{62}   & \gradientcell{64}     & \gradientcell{9}   & \gradientcell{6}    & \gradientcell{24} & \gradientcell{27}  \\
DNSMOS (SIG)&   & \gradientcell{26} & \gradientcell{26}   & \gradientcell{48} & \gradientcell{49}    &  \gradientcell{55} & \gradientcell{53}  %
                             & \gradientcell{77}   & \gradientcell{84}&   \gradientcell{85} & \gradientcell{78}    & \gradientcell{85} & \gradientcell{84} %
                         & \gradientcell{41}   & \gradientcell{28} & \gradientcell{6} & \gradientcell{9}   & \gradientcell{19} & \gradientcell{12}  \\                         
DNSMOS (OVRL)&     & \gradientcell{35} & \gradientcell{35}   & \gradientcell{61} & \gradientcell{56}   &  \gradientcell{66} & \gradientcell{60}  %
                                & \gradientcell{90}   & \gradientcell{95} & \gradientcell{83} & \gradientcell{83}   & \gradientcell{93} & \gradientcell{95} %
                         & \gradientcell{51}   & \gradientcell{47} & \gradientcell{5} & \gradientcell{6}    & \gradientcell{23} & \gradientcell{22}  \\
                         \midrule
\multicolumn{20}{l}{Audiodec$_\textrm{libriTTS}$}\\
\midrule          
$\overline{\textrm{LQR}}$  &    & \gradientcell{55}  & \gradientcell{47}      & \gradientcell{89} & \gradientcell{88}       &  \gradientcell{90} & \gradientcell{91}  %
                                & \gradientcell{99}  & \gradientcell{97}      & \gradientcell{75} & \gradientcell{85}       & \gradientcell{97} & \gradientcell{97} %
                                & \gradientcell{70} & \gradientcell{68}    &   \gradientcell{10} & \gradientcell{9}   & \gradientcell{39} & \gradientcell{39}  \\
$\textrm{LQR}_{0, K}$       &   & \gradientcell{59} & \gradientcell{55}  & \gradientcell{90} & \gradientcell{87}     &  \gradientcell{90} & \gradientcell{90}  %
                                & \gradientcell{99}   & \gradientcell{97}  & \gradientcell{75} & \gradientcell{85}   & \gradientcell{98} & \gradientcell{98} %
                                & \gradientcell{69}   & \gradientcell{68}   & \gradientcell{10} & \gradientcell{8}     & \gradientcell{39} & \gradientcell{39}  \\
\midrule
\multicolumn{20}{l}{Encodec$_{1.5}$}\\
\midrule    
$\overline{\textrm{LQR}}$ &     & \gradientcell{57}      & \gradientcell{47} & \gradientcell{90} & \gradientcell{92}         &  \gradientcell{87} & \gradientcell{91}  %
                                & \gradientcell{95}      & \gradientcell{96} & \gradientcell{56} & \gradientcell{86}         &   \gradientcell{88}  & \gradientcell{96} %
                                & \gradientcell{67}      & \gradientcell{66} & \gradientcell{1} & \gradientcell{1}      &   \gradientcell{30}  & \gradientcell{31}  \\
$\textrm{LQR}_{0, K}$    &      & \gradientcell{56}      & \gradientcell{44} & \gradientcell{89} & \gradientcell{91}     &   \gradientcell{85} & \gradientcell{90}  %
                                & \gradientcell{95}      & \gradientcell{96} & \gradientcell{56} & \gradientcell{86}     &   \gradientcell{88} & \gradientcell{96} %
                                & \gradientcell{68}      & \gradientcell{67} & \gradientcell{1} & \gradientcell{9}      &   \gradientcell{30} & \gradientcell{31}  \\
\midrule
\multicolumn{20}{l}{Funcodec General}\\
\midrule   
$\overline{\textrm{LQR}}$ &     & \gradientcell{15}      & \gradientcell{17}      & \gradientcell{14} & \gradientcell{23}     & \gradientcell{17}  & \gradientcell{24} %
                                & \gradientcell{86}      & \gradientcell{67}      & \gradientcell{36} & \gradientcell{49}     & \gradientcell{74} & \gradientcell{66}  %
                                & \gradientcell{63}      & \gradientcell{64}      & \gradientcell{6} & \gradientcell{9}       & \gradientcell{22}  & \gradientcell{22}  \\
$\textrm{LQR}_{0, K}$    &      & \gradientcell{18}      & \gradientcell{19}    & \gradientcell{22} & \gradientcell{27}      & \gradientcell{25} & \gradientcell{29}%
                                & \gradientcell{88}      & \gradientcell{69}    & \gradientcell{39} & \gradientcell{53}     & \gradientcell{76}  & \gradientcell{68}  %
                                & \gradientcell{66}      & \gradientcell{67}    & \gradientcell{3} & \gradientcell{6}       & \gradientcell{25} & \gradientcell{25}  \\
\midrule
\multicolumn{20}{l}{Funcodec Academic}\\
\midrule   
$\overline{\textrm{LQR}}$ &     & \gradientcell{24}     & \gradientcell{27} & \gradientcell{35} & \gradientcell{40}      & \gradientcell{38}  & \gradientcell{41} %
                                & \gradientcell{84}     & \gradientcell{67} & \gradientcell{34} & \gradientcell{50}       &  \gradientcell{72} & \gradientcell{66}  %
                                & \gradientcell{64}     & \gradientcell{66}& \gradientcell{3} & \gradientcell{6}       & \gradientcell{23}  & \gradientcell{25}  \\
$\textrm{LQR}_{0, K}$    &      & \gradientcell{21}     & \gradientcell{13} &  \gradientcell{33} & \gradientcell{33}        & \gradientcell{38} & \gradientcell{32} %
                                & \gradientcell{85}     & \gradientcell{65}& \gradientcell{36} & \gradientcell{46}      &  \gradientcell{74} & \gradientcell{63}  %
                                & \gradientcell{67}     & \gradientcell{71} &  \gradientcell{1} & \gradientcell{6}         & \gradientcell{26} & \gradientcell{28}  \\
\bottomrule
    \end{tabular}
    \label{tab:UDASE_table}
    \vspace{-4mm}
\end{table*}

\vspace{-2mm}
\subsection{Results on CHiME-7 UDASE}
Table \ref{tab:UDASE_table} shows the results for the training session and the in-domain unlabeled eval set of CHiME-7 UDASE. Intrusive measures can be computed only for the training session. 

As expected, a weak correlation is observed in all cases when considering perceived BAK on the SIG-only signals. This is because no additive background noise is present in this subset, making the ground-truth data collapse on the best BAK scores, for which small variations in scores are simply due to noise in the subjective measurement, and are not expected to be reflected in the objective metrics.

A strong to very strong correlation is observed for all intrusive metrics when predicting SIG and OVRL on the SIG-only subset. In contrast, the non-intrusive metrics from the literature generally exhibit a weaker correlation.
Nonetheless, the non-intrusive LQR is shown to match the performance of the intrusive metrics when considering Audiodec$_\textrm{libriTTS}$ and Encodec$_{1.5}$, which were also among the best cases in Table~\ref{tab:ODAQ_CC}.

When considering BAK only, we would have expected a weak to very weak correlation with the perceived SIG, similar to the case of perceived BAK for the SIG-only subset. Inspecting the ground-truth data reveals that rated SIG is not collapsed to the best scores as expected, but a significant portion of the MOS range is covered, i.e., $[3.4, 5.0]$. A possible explanation is the complexity of the task for the listeners to ignore the background noise. At least at very low SNR levels, listeners might perceive the additive noise as affecting the speech quality even though the speech signals are not distorted but simply presented in the presence of noise.

This BAK-only subset is particularly easily predicted by the objective metrics, from energy-based SIR to more complex metrics such as LQR.
Also in this case, Audiodec$_\textrm{libriTTS}$ and Encodec$_{1.5}$ show the best performance.

Considering the in-domain unlabeled eval, LQR often matches or outperforms the non-intrusive metrics from the literature when looking at the correlation with perceived BAK.
A very weak correlation is observed for perceived SIG and OVRL for all objective measures, as also noted in \cite{LEGLAIVE2025}.
A possible explanation is that, in this listening test, differences in terms of perceived BAK are more marked than the ones perceived for SIG and OVRL. The per-system mean BAK spans a range of $2.10$ points on the 5-point rating scale. For SIG, the spanned range is only $1.34$ points and $0.97$ for OVRL, i.e., smaller than the scoring resolution available to the listeners, which was $1.0$ in this listening test.
The finer granularity of the quality differences and the coarse rating resolution pose additional challenges in terms of precision for the subjective ground-truth scores and the objective metrics.

In all cases, lower performance is observed for Funcodec. We hypothesize that structured quantization dropout decreases the dependency between the signal quality and its discrete latent representation. Further investigations are essential to verify this hypothesis. 

\vspace{-0mm}
\section{Conclusions and Final Remarks}
In this paper, the relation between the speech signal model learned by a neural codec and the perceived quality of an input speech signal is investigated. In particular, to quantify this relation, \ac{LQR} is introduced as a proxy for the distance or deviation from the learned speech signal model for a given input signal. As shown by the results, \ac{LQR}-based metrics exhibit a strong correlation with subjective speech quality that is comparable to, or larger than, that of other intrusive and non-intrusive metrics. This motivates viewing \ac{LQR} as the basis for more sophisticated speech quality metrics, which, unlike the \ac{LQR}-based metrics in this paper, can be trained specifically to predict speech signal quality. 


\clearpage
\newpage

\bibliographystyle{IEEEtran}
\bibliography{lib}

\end{document}